
\input jydoc
\typesize=12pt

\jyinput article

\def\du_#1^#2{{\scriptstyle\p3}\llap{$_{#1}$}^{#2}}
\def\ud^#1_#2{{\scriptstyle\p3}\llap{$^{#1}$}_{#2}}

\def\om{\omega}
\def\alb{\overline{\alpha}_S}
\def\lamb{\bar{\lambda}}

\equation{1.1}{
M(\om,r) = \int_0^1 dx_B \, x_B^{N-1} \, x_BG(x_B,q^2) = 
\int_0^\infty dy \, e^{-\om y} [x_B G(x_B,q^2)] 
,}

\Equation{1.2}{\eqalignno{
M(\om,r) = C_2(\om,r) \langle p |O^{(2)}|p\rangle
+ & {1\over Q^2} C_4(\om,r) \langle p |O^{(4)}|p\rangle + \ldots \cr
+ &{1\over Q^{2i-2}} C_{2i}(\om,r) \langle p |O^{(2i)}|p\rangle
+ \ldots, &\eqnlabel{1.2}
}}

\equation{1.3}{
C_{2i}(\om,r)  \sim e^{\gamma_{2i}(\om)r} 
,}

\equation{1.4}{
\gamma_2(\om) = {N_c \alpha_S \over \pi\om}
. }

\equation{1.5}{
\gamma_{2n}(\om) = n\gamma_2({\om \over n}) 
.}

\equation{1.6}{
C_4(\om,r) \langle p |O^{(4)}|p\rangle = \int {d\om' \over 2\pi i}
C_2(\om',r) C_2(\om-\om') \langle p |O^{(4)}|p\rangle 
\sim \int d\om' e^{\gamma_2(\om-\om')r + \gamma_2(\om')r}
. }

\equation{1.7}{
\gamma_4(\om) = 2 \gamma_2({\om \over 2}) [1+\delta]=
{4 N_c \alpha_S \over \pi\om} [1+\delta] 
,}

\equation{1.8}{
\gamma_2(\om_{cr}) = -1 +  \gamma_4(\om_{cr}) 
.}


\equation{2.1.1}{
P(z) = {N_c \alpha_S \over \pi z} \equiv {\alb \over z}
.}

\equation{2.1.2}{
F(x,\hat{q}_{1t}^2 = \hat{q}_{2t}^2) = \alb \int^1_{x}{dz\over z}
\int_0^{\ln\hat{q}_1^2=\ln\hat{q}_2^2} dr
F({x\over z},r)
.}

\equation{2.1.3}{
F(\om,f) = \int_0^\infty dr \, e^{-fr} F(\om,r)
.}

\equation{2.1.4}{
\om f = \alb
.}

\equation{2.1.5}{
G_2(\om,f) = {1 \over \om f - \alb}
.}

\equation{2.1.6}{
F(x,\hat{q}_{1t}^2,\hat{q}_{2t}^2) = {(q_{0,1t}\cdot q_{0,2t}) \over
q_{0,1t}^2} F(x,\hat{q}_{2t}^2)
}

\Equation{2.2.1}{\eqalignno{
C_4(\om,r=\ln(Q^2))&\langle p |O^{(4)}|p\rangle\;\Big|_{\rm 2P}  = 
\int {d\om' \over 2\pi i} M(\om',r)M(\om-\om',r) <S> \cr
= & \int df e^{fr} G_2(\om',f')G_2(\om-\om',f-f') {d\om'df' \over 2\pi i}
<S>. & \eqnlabel{2.2.1}
}}

\equation{2.2.2}{
C_4(\om,r)\langle p |O^{(4)}|p\rangle\;\Big|_{\rm 2P}
 = \int {df\over 2\pi i}
{e^{fr} \over \om f \sqrt{1-{4\alb \over \om f}}} <S>
.}

\equation{2.2.3}{
\gamma_4 = {4\alb \over \omega}
.}

\equation{saddlef1}{
C_4(y,r) \approx \int {df df' \over f' (f-f')} e^{fr} e^{({1\over f'} +
{1\over f-f'})\alb y} 
.}

\equation{saddlef2}{
f' = {f \over n}
.}

\equation{saddlef3}{
C_4(y,r) \sim \int d\om \;\; e^{{4\alb \over \om}r+\om y}
,}

\Equation{2.3.1}{\eqalignno{
 C_4(Y,Q^2) & \langle p |O^{(4)}|p\rangle\; \Big|_{\rm PI} = 
&\eqnlabel{2.3.1}\cr
 {1 \over 2}{1 \over N_c^2-1} 2 & \alb \int^Y_0 dy_1 
\int_{l_{0,1}^2}^{Q^2}{dl_1^2\over l_1^2}
F(Y-y_1,r_Q - r_{l_1})^2 \cr
\times & \Big\{ \int^{l_1^2}_{l_{0,1}^2} {dl_1'^2\over l_1'^2}
\int^{y_1}_0 dy_1' \tilde{F}(y_1', r_{l_1'}) 
\int^{l_1^2}_{l_{0,2}^2}{dl_2^2\over l_2^2}
\int^{y_1}_0 dy_2  \tilde{F}(y_2, r_{l_2})  \cr
 + & \int^{y_1}_0 dy_2 \int^{y_2}_0 dy_1' 
\int^{l_1^2}_{l_{0,1}^2} {dl_1'^2\over l_1'^2}
\tilde{F}(y_1', r_{l_1'})
\int^{l_1^2}_{l_{0,2}^2}{dl_2^2\over l_2^2}
\int^{y_1}_0 dy_2  \tilde{F}(y_2, r_{l_2})\Big\} 
}}

\equation{2.3.2}{
C_4(\om,r)\langle p |O^{(4)}|p\rangle\Big|_{\rm PI} = 
\delta \cdot \int {e^{fr} df\over 2\pi i} {1\over \om f
 \sqrt{1-{4\alb \over \om f}}}
\{ {1 \over \sqrt{1-{4\alb \over \om f}}} -1 \} <S>
}

\equation{2.4.1}{
C_4(\om,r)\langle p |O^{(4)}|p\rangle = 
\int {{e^{fr}} df \over 2\pi i} {1\over \om f \sqrt{1-{4\alb \over \om f}}
(1- \delta[{1\over \sqrt{1-{4\alb \over \om f}}} -1])}<S>
}

\equation{2.4.2}{
1- \delta[{1\over\sqrt{1-{4\alb \over \om f}}} -1] = 0, \quad 
f_0 = {4\alb \over \om}
(1+\delta^2)
.}

\equation{2.4.3}{
{2\delta\over \om f - 4\alb(1-\delta^2)}.
}

\equation{2.4.4}{
\gamma_4 = {4\alb(1+\delta^2) \over \om}
}

\equation{2.4.5}{
 \lambda \int {e^{fr}df\over 2\pi i}
{1\over (\om f \sqrt{1-{4\alb \over \om f}})^2}
}

\equation{2.4.6}{
{1\over \om f \sqrt{1-{4\alb \over \om f}} - \lambda }
}

\equation{3.1}{
(\om - \sum_k {\alb\over f_k})^{-1}
.}

\equation{3.2}{
(1+i\Delta_k){f\over n} = f_k
.}

\equation{prop2}{
(\om -  {\alb n^2 \over f} +{\alb n\over f}\sum_k^n\Delta_k^2)^{-1}
.}

\equation{newenergy}{
{\cal E}_n = {-\om +(\alb/f)n^2 \over (\alpha_S n/f)}
}

\equation{Hammom}{
H = {1\over 2m}\sum_k \Delta_k^2 a_k^{\dagger} a_k
 + \lambda \sum_k a_k^{\dagger} a_k^{\dagger} a_k a_k
,}

\equation{Hamcoo}{
H = \int dx \, \partial_x \psi^{\dagger}\partial_x \psi
+ \bar{\lambda} : \psi^{\dagger}\psi^{\dagger}\psi\psi:
, \quad \bar{\lambda} = {\lambda \over 2m}. }

\equation{numbop}{
N = \int dx \, : \psi^{\dagger}(x) \psi(x):
.}

\equation{eigen}{
H |\psi_n\rangle = {\cal E}_n |\psi_n\rangle \;,\;
N|\psi_n\rangle = n |\psi_n\rangle
.}

\equation{eigfun}{
|\Psi_n\rangle = {1\over \sqrt{n!}}\int d^nz
\chi_n(z_1,\ldots,z_n) \psi^{\dagger}(z_1)\ldots\psi^{\dagger}(z_n)
| 0\rangle
,}

\equation{Fockdef}{
\psi(x)| 0\rangle=0 \; ; \; \langle 0| \psi^{\dagger}(x)=0\; ; \;
\langle 0 | 0 \rangle = 1
}

\equation{eigenz}{
{\cal H}_n \chi_n = {\cal E}_n \chi_n 
,}

\equation{Hamz}{
{\cal H}_n = -\sum_{j=1}^n {\partial^2\over \partial z_j^2}
+ 2\bar{\lambda}\sum_{n\geq k > j \geq 1} \delta(z_k-z_j)
.}

\equation{Hamfree}{
{\cal H}^0_n = -\sum_{j=1}^n {\partial^2\over \partial z_j^2} \;\;,\;\;
{\cal H}^0_n \chi^0_n = {\cal E}_n \chi^0_n
}

\equation{bound}{
({\partial \over \partial z_{j+1}}-{\partial \over \partial z_j})
\chi_n = \lamb \chi_n \;\;,\;\; (z_{j+1}=z_j+0, \; j=1\ldots n)
.}

\equation{chi0}{
\chi^0_n = \det e^{i\kappa_j z_k}
,}

\equation{enfree}{
{\cal E}_n = -\sum_{j=1}^n \kappa_j^2
.}

\equation{chiint}{
\chi_n = {\rm Const}\, \Pi_{j>k}
({\partial \over \partial z_{j+1}}-{\partial \over \partial z_j}
+\lamb)\det e^{i\kappa_j z_k}
.}

\equation{detP}{
\det e^{i\kappa_j z_k} = \sum_{\rm P} (-)^{[\rm P]}
e^{i\sum_n \kappa_{{\rm P}_k} z_k}
,}

\equation{chifin}{
\chi_n = {\rm Const}\sum_{\rm P} (-)^{[\rm P]}\Pi_{j>k}
(\kappa_{{\rm P}_j}-\kappa_{{\rm P}_k}-
i\lamb\varepsilon(z_j-z_k))e^{i\sum_n \kappa_{{\rm P}_k} z_k}
}

\equation{sumlam}{
\sum_{i=1}^n \kappa_i = 0
.}

\Equation{lambdas}{\eqalignno{
\kappa_k^n = & ({2k-n-1 \over 2}) i\lamb \;\;\;\quad \cr
 k & = 1,2,\ldots,n & \eqnlabel{lambdas}
}}

\equation{enint}{
{\cal E}_n = \sum_{j=1}^n {\cal E}^j_n =
-\sum_{j=1}^n (\kappa_j^n)^2 = 
{\lamb^2 \over 12} (n^3-n)
.}

\equation{gamma2n}{
\gamma_{2n} = {\alb n^2\over \om} (1+{\delta^2 \over 3}(n^2-1))
,}

\equation{chin=2}{
\chi_2 = (\kappa_2-\kappa_1-i\lamb) e^{i(\kappa_1 z_1 +
\kappa_2 z_2)} - (\kappa_1-\kappa_2-i\lamb)
e^{i(\kappa_2 z_1 +\kappa_1 z_2)} \;,\; z_1<z_2
.}

\equation{chin=2fin}{
\chi_2 = {\lamb\over \sqrt{2|\lamb|}}\; e^{-|\lamb|
|z_2-z_1|/2}
.}

\equation{chin=2p}{
\chi_2 = (\kappa_2-\kappa_1+i\lamb) e^{i\kappa_1(z_2-z_1)}
 - (\kappa_1-\kappa_2+i\lamb)
e^{-i\kappa_1(z_2 -z_1)} \;,\; z_1<z_2
.}

\equation{gamma4p}{
\gamma_4(\om) = {4\alb\over \om}(1+\delta^2)
.}

\Equation{chin=3}{\eqalignno{
\chi_3 \quad  \propto\quad & e^{i x_3\kappa_{31}+ i x_2\kappa_{21}}
	(\kappa_{32}-i\lamb)(\kappa_{31}-i\lamb)(\kappa_{21}-i\lamb) \cr
        & + e^{-i x_3\kappa_{21}+ i x_2\kappa_{32}}
	(\kappa_{31}+i\lamb)(\kappa_{21}+i\lamb)(\kappa_{32}-i\lamb) \cr
        & + e^{-i x_3\kappa_{32}-i x_2\kappa_{31}}
	(\kappa_{21}-i\lamb)(\kappa_{32}+i\lamb)(\kappa_{31}+i\lamb) \cr
        & + e^{i x_3\kappa_{21}+ ix_2\kappa_{31}}
	(\kappa_{32}+i\lamb)(\kappa_{21}-i\lamb)(\kappa_{31}-i\lamb) \cr
        & + e^{i x_3\kappa_{32}-i x_2\kappa_{21}}
	(\kappa_{31}-i\lamb)(\kappa_{21}-i\lamb)(\kappa_{21}+i\lamb) \cr
        & + e^{-i x_3\kappa_{31}-i x_2\kappa_{32}}
	(\kappa_{21}+i\lamb)(\kappa_{31}+i\lamb)(\kappa_{32}+i\lamb) 
	& \eqnlabel{chin=3}
}}

\equation{chi3bound}{
\chi_3 = \lamb\, e^{2|x_3|\lamb
+|x_2|\lamb}\quad,\quad\lamb<0
.}

\equation{chiagain}{
\chi_n = {\rm Const}\sum_{\rm P} (-)^{[\rm P]}\Pi_{j>k}
(\lambda_{{\rm P}_j}-\lambda_{{\rm P}_k}-
i\lamb\varepsilon(z_j-z_k))e^{i\sum_n \lambda_{{\rm P}_k} z_k}
}

\equation{App1}{
F(y,r) = \int {d\om'df' \over (2\pi i)^2} 
G_2(\om',f') e^{\om'y + f' r} ,
}

\equation{App2}{
{\alb \over (N_c^2-1)}{1 \over \om f 
\sqrt{1-{4\alb \over \om f}}}.
}

\Equation{App3}{\eqalignno{
& \int_{-\infty}^{\infty}dy_1 \int_{-\infty}^{\infty}d r_{l_1}
e^{-\om y_1 -f r_{l_1}}
\int {d\om_1 df_1 \over (2\pi i)^2}
\int {d\om_2 df_2 \over (2\pi i)^2} \cr
& (1+{\om_2 \over \om_1}) {e^{\om_1 y_1 + f_1 r_{l_1}}
\over \om_1 f_1 -\alb} {e^{\om_2 y_1 + f_2 r_{l_1}}
\over \om_2 f_2 -\alb}
& \eqnlabel{App3}
.}}

\equation{App4}{
\int {df_2\over 2\pi i} {\om f_2\over \om f_2 - \alb}
{1\over \om f_2(f - f_2) -\alb f}
.}

\equation{App5}{
{1 \over 2} {1\over \alb}\Big( {1 \over 
\sqrt{1-{4\alb \over \om f}}}-1 \Big).
}

%
%
%
%
\pagenumstyle{blank}
\footnotenumstyle{arabic}

\line{August 1993\hfil FERMILAB-PUB-93/243-T}

\vskip7em

\centertext{\bigsize\bfs  Anomalous Dimensions of High Twist Operators in QCD\\
	at $N \rightarrow 1$ and large $Q^2$}

\vfil

\begin{center}

E. Laenen\footnote{e-mail: eric@fnth010.fnal.gov}, E. Levin\footnote{on
leave from St. Petersburg Nuclear Physics Institute, 
188350 Gatchina, St. Petersburg, Russia\\
e-mail: levin@fnal.fnal.gov, FNALV::LEVIN}

\medskip

{\it Fermi National Accelerator Laboratory\\
P.O.Box 500, MS 106, Batavia, Illinois 60510}

\bigskip

A.G. Shuvaev\footnote{e-mail: shuvaev@lnpi.spb.su, galinast@lnpi.sbp.su}

\medskip

{\it St. Petersburg Nuclear Physics Institute\\
188350 Gatchina, St. Petersburg, Russia}

\end{center}

\vfil

\centertext{\bfs Abstract}
\vskip\belowsectionskip

\begin{narrow}[4em]

The anomalous dimensions of high-twist operators
in deeply inelastic scattering  ($\gamma_{2n}$)
are calculated in the limit when
the moment variable $N \rightarrow 1$ 
(or $x_B\rightarrow 0$) and at large $Q^2$ (the double logarithmic
approximation) in perturbative QCD. We find that the
value of $\gamma_{2n}(N-1)$ 
in this approximation  behaves as
${N_c \alpha_S \over \pi (N-1)} n^2(1 + {\delta \over 3}
(n^2-1))$ where $\delta \approx 10^{-2}$.
This implies that the contributions 
of the high-twist operators give rise to an earlier onset of shadowing
than was estimated before.
The derivation makes use of a Pomeron exchange approximation,
with the Pomerons interacting attractively. 
We find that they
behave as a system of fermions.

\end{narrow}

\vfil

\vskip1em

\break


\sectionstyle{left}
\sectionnumstyle{arabic}
\pagenumstyle{arabic}
\pagenum=0
\footnotenumstyle{arabic}
\footnotenum=0

\null
\vskip0pt plus.5\baselineskip

{\bfs\section{Introduction}}

\vskip0pt plus.5\baselineskip

To convey the main goal and result of this 
paper we start by recalling the principal 
steps of the theoretical approach to deep-inelastic
scattering.
We first introduce the moments of the deep-inelastic structure
function, 
$$\putequation{1.1}$$
where $\omega=N-1$, $y=\ln(1/x_B)$ and $r=\ln(q^2/q_0^2)$, 
$x_B$ denoting the Bjorken scaling variable, and $q_0^2$ 
an IR cut-off.
Each moment is given through a  Wilson Operator Product
Expansion (OPE) in the form
$$\putequation{1.2}$$
where $C_i$ is the coefficient function and 
$\langle p|O^{(2i)}|p\rangle$ denotes generically 
the expectation value of a 
twist $2i$ operator in a proton state
(see [\putref{Coll}] for details).
In practice, one usually neglects 
all high twist contributions (i.e. all terms in
(\puteqn{1.2}) beyond the first), by assuming that they are
all supressed at large values of $Q^2$ due to the factor
of $Q^{-2i}$ in front.
It is well known from renormalization group arguments that a 
coefficient function $C_{2i}$ behaves as 
$$\putequation{1.3}$$
where $\gamma_{2i}$ is the anomalous dimension of the twist $2i$ operator
\footnote{for simplicity we consider here the case of a fixed
$\alpha_S$.}.
The anomalous dimension of the leading twist contribution can
be calculated using the Gribov-Lipatov-Altarelli-Parisi (GLAP)
equation and is equal to 
$$\putequation{1.4}$$
The specific contribution
to the anomalous dimensions
of high-twist operators originating from the exchange 
of $n$ `leading twist ladders' in the t-channel
was found in the GLR paper [\putref{GLR}]. The result was
$$\putequation{1.5}$$
Briefly, to illustrate the above statement, consider the twist four
contribution. The two-ladder exchange leads to the following
contribution in this case
$$\putequation{1.6}$$
This integral has a saddle point at $\omega' = \omega/2$ so 
$C_4 \sim \exp(2\gamma_2({\omega \over 2})r)$. Thus
$\gamma_4 = 2\gamma_2({\omega \over 2})r$.

Recently Bartels on the one hand [\putref{B}]
 and Levin, Ryskin and Shuvaev on the other [\putref{LRS}]
performed the next
step in understanding the high twist contribution to (\puteqn{1.1}).
Both groups calculated the anomalous dimension of the twist
four gluon operator, each using quite different techniques. The 
value of the anomalous dimension was found to be
$$\putequation{1.7}$$
with $\delta ={\cal O}((N_c^2-1)^{-2}) \approx 10^{-2}$ small.

The most important lesson to learn from this calculation is the fact that one
{\it  cannot} trust the GLAP evolution equation in the region
of small $\omega$ (or, equivalently, large $\ln(1/x_B)$). Indeed for
$\omega$ smaller than some $\omega_{cr}$ the twist four contribution 
becomes
larger than the leading twist one. The value of $\omega_{cr}$
can be found from the equation
$$\putequation{1.8}$$
The same conclusion could be arrived at using the GLR
approach but in [\putref{B},\putref{LRS}] 
this statement was proved for the 
whole set of Feynman diagrams instead of just the two-ladder
contribution that the GLR approach takes into account.

In this paper we will calculate the
anomalous dimension for an arbitrary high twist
operator, beyond the result (\puteqn{1.5}). 
The operators in question are so called gluonic Quasi-Partonic Operators
(QPO's), which were introduced in [\putref{EFP}] and 
have been studied in detail in [\putref{BFL}].
Some of their properties relevant to this paper
are the following:
1) the twist of these operators coincides with the number
of gluonic fields they contain; 2) under a scale change 
a QPO can only
transform into a QPO or in operators that can be
expressed in terms of QPO's using equations of motion.
The transformation from other operators into
QPO's is possible, but not vice versa. Thus, such
a transformation cannot change the value of the
anomalous dimension; 3) the evolution 
equation for such operators looks like
a Faddeev-type equation with a pair-like interaction
(with the number of particles conserved),
the kernels of which are the same as for twist two
operators. 
The analogy with the Faddeev-equation is very 
important for us since
the method by which we will arrive at these anomalous dimensions
has also been used in the nonrelativistic
three body problem (for which the Faddeev equation
was originally derived): 
we try to extract the resonance-like
interaction in the two gluon channel first and
subsequently take into account the interaction between
such `resonances'. The so-called Pomeron (colorless ladder
for two gluon exchange in the t-channel) plays the r\^ole
of such a resonance in our calculation.
In refs. [\putref{B},\putref{LRS}]
it was shown that the value of the twist four anomalous dimension
in the double logarithmic approximation 
is determined mainly by the exchange, and interaction, of two
colourless gluon-`ladders' (Pomerons) in the t-channel,
and that the interaction
between such ladders is small since it
is proportional to  $1/(N_c^2 - 1)$.
This indicates that our method of 
taking into account the two gluon 
interaction first, creating a Pomeron, is right,
since it is the correct approach to the 
problem at least as $N_c \rightarrow \infty$. 
The solution to the problem at finite $N_c$ we present
in this paper. 
The above observations considerably simplify the problem
and will enable us to reduce it to solving the Nonlinear Schrodinger
Equation for $n$ Pomerons in the t-channel.

The paper is organized as follows: in section 2 we consider
in detail the calculation of the twist four anomalous dimension
in the two-Pomeron approximation. We can find the value of 
$\gamma_4$ by `summing' all diagrams in an explicit way and we
will use this concrete example to introduce all notations and
to illustrate all further steps in finding the
value of $\gamma_{2n}(\omega)$. In section 3 we reinterpret the calculation
of $\gamma_{2n}(\omega)$ in terms of a two-dimensional theory 
describing the interaction of $n$ nonrelativistic particles. 
In section 4 we find the energy of the ground state
of this theory, which corresponds precisely to $\gamma_{2n}(\omega)$.
In the conclusions we summarize our results and discuss
outstanding problems. 
In Appendix A we discuss some technical details.

{\bfs\section{The anomalous dimension of the twist 4
 operator in the two-Pomeron approximation}}

In this section we will derive the anomalous dimension of the
twist 4 operator from diagrams with four gluon exchange in 
the t-channel. Our method consists of first pairing 
the gluons up into `Pomerons', and subsequently calculate
the interaction between the latter. 

The contributions to the anomalous dimensions of the twist
four operator due to Pomeron exchange can be separated into
two cases: in one the Pomerons do not interact 
while being exchanged, in the other they do. We will discuss these
cases in turn, but begin by discussing just the two-gluon (one
Pomeron) exchange to set the stage. In addition, this exchange
is a building block for the four gluon case.

\subsection{2.1.\ {\it The Pomeron in the double log approximation} (DLA).}

We start from the structure function of a two-gluon colorless
state (Pomeron) in the DLA of perturbative QCD (pQCD), see Fig.1.
We will generically denote this function by $F(x,q^2)$. 
This is a step-up to our discussion of the four-gluon state
later on in this section.
We discuss this function for two cases, distinguished by the variable
$q_t$, the transverse momentum along
the ladder. In the first $q_t = q_{1t} - q_{2t}
= q_{0,1t}-q_{0,2t} = 0$ (see Fig.1 for notation). 
The two-gluon structure function
can be calculated directly from the GLAP [\putref{GL}] evolution
equation, with kernel
$$\putequation{2.1.1}$$
where we have dropped terms that are non-singular as $z\rightarrow 0$.
In DLA the GLAP evolution equation can be written in the
form
$$\putequation{2.1.2}$$
where the hat indicates that 
the momentum has been divided by the corresponding
lower cut-off momentum, e.g. $\hat{q}_{1t} \equiv q_{1t}/q_{0,1t}$, etc.
The choice of IR cut-off of the 
GLAP evolution equation is arbitrary. In Fig.1
we depicted the cut-off to lie somewhere along
the ladder, but we could have also chosen it be 
close to the `blob', in which
case we would have
$q_{0,1t} \simeq q_{0,2t} \simeq 1/R_h$, very small.
Eq. (\puteqn{2.1.2}) can be simplified further
by performing two Mellin transforms, one with respect to
$y = \ln(1/x)$, as in (\puteqn{1.1}), the other 
with respect to $r$, 
$$\putequation{2.1.3}$$
In $\omega,f$ representation eq. (\puteqn{2.1.2}) has the
form
$$\putequation{2.1.4}$$
We define $G_2(\omega,f)$ by
$$\putequation{2.1.5}$$
It gives the Green function
of the GLAP equation in DLA, which satisfies
$G_2(x,q_{t}^2) = \theta\big(\ln(\hat{q}_t^2)\big)$ at $x=1$. 

In the second case we have $q_t \neq 0$,
in other words momentum is transferred along the ladder. 
We will encounter this situation a little further on.
Within DLA, the contributions come from two regimes:
1) $q_t^2 << q_{1t}^2 \approx q_{2t}^2$
and 2) $q_t^2 \approx q_{0,1t}^2 >> q_{0,2t}^2$, with
$q_{0,2t}^2 \simeq 1/R_h^2$. 
In spite of the fact that now $q_{1t} \neq q_{2t}$ 
the two gluon structure function in essence still depends 
only on one virtuality in these regimes, viz. the smallest one.
E.g. for $|q_{2t}| < |q_{1t}|$, we have, for both regimes
$$\putequation{2.1.6}$$
For values of $q_t$ other than the ones discussed in this
subsection one does not find a large logarithmic contribution.
The derivation of and all details related to eq. (\puteqn{2.1.6}) 
can be found in refs. [\putref{GLR,B,LRS,GL}]. We now go on
to discuss the four-gluon structure function.

\subsection{2.2.\ {\it Two-Pomeron case} (2P).}

Let us consider now the simplest diagram that gives a contribution
to the four-gluon structure function and thus to 
$\gamma_4$, namely the exchange of two `leading twist' ladders 
(see Fig.2).
In DLA we can neglect the dependence of the deep-inelastic scattering
structure function on the momentum $q_t$
(see [\putref{GLR}] for the relevant discussion), and the contribution
of this diagram can be written in the form 
$$\putequation{2.2.1}$$
where $<S>$ 
stands for the integration over $q_t$ of the
matrix element that describes the emission of 4 gluons from
the nucleon. 
Substituting eq. (\puteqn{2.1.5}) in (\puteqn{2.2.1}) and performing
the integrals over $\omega'$ and $f'$ by contour integration we get
$$\putequation{2.2.2}$$
The contour of integration over $f$ is located to the right
of all singularities in this variable.
From (\puteqn{2.2.2}) one can easily see that the main 
contribution comes from $f\rightarrow {4\alb \over \omega}$
and it is proportional to $\exp({4\alb \over \omega}r)$.
Thus the contribution to the value of the anomalous dimension 
from the diagram in Fig.2 is 
$$\putequation{2.2.3}$$
This will turn out to be the most significant contribution
to $\gamma_4$.

An alternative method of derivation consists of using the
saddle point method for the $f'$ integral, as was done for 
the $\om'$ integration in (\puteqn{1.6}). 
Starting with (\puteqn{2.2.1}), applying an inverse
Mellin transform with respect to $\omega$ and performing 
the $\om, \om'$ integrals by contour integration one
obtains
$$\putequation{saddlef1}$$
One can now do the $f'$ by the saddle point method, yielding
the saddle point $f' = f/2$. For $n$ ladders one would obtain
$$\putequation{saddlef2}$$
Changing variables $f \rightarrow 4\alb/\om$ then
yields
$$\putequation{saddlef3}$$
and thus we find again $\gamma_4 = 4\alb/\om$.

\subsection{2.3.\ {\it Pomeron interaction} (PI).}

However, although (\puteqn{2.2.3}) is the most significant contribution
to $\gamma_4$, it is not the full answer. As was shown
in ref. [\putref{B},\putref{LRS}] the Pomeron-Pomeron interaction 
crucially changes the value of the anomalous dimension. In this subsection we 
want to understand 
this statement, and to that end let us consider the diagram in Fig.3, which
displays this interaction. We will discuss here the contribution to
$C_4 <p|O^{(4)}|p>$  from this diagram. All notation is 
explained in the figure, where all $y$'s are rapidities
and all other symbols are
transverse momenta. (E.g. $k_i$ is the transverse momentum 
at rung $i$ in the ladder in the left upper part of the
diagram.)

Clearly, there are two distinct parts to 
this two ladder diagram:  the top and bottom half, and we will
discuss the contribution from this diagram accordingly.
The obvious distinction from Fig.2 is that the two 
ladders in the top half are `interchanged' in the bottom half.
The main contribution from this diagram in DLA is
when all momenta in the top half
(above the dashed line)
of the diagram are larger than all momenta in the
bottom half, yet smaller than $Q^2$. Note that while
there is momentum transferred along the ladders in 
the top half, that is not the case in the bottom
half.

As far as the momenta in the bottom half are concerned,
we assume
$l_1^2 >> l_2^2$. The case
$l_2^2 >> l_1^2$ gives precisely the same answer, 
so we only discuss the former.
The IR cutoff momenta $l_{0,1},l_{0,2}$ are small,
$O(1/R_h)$. Each individual (piece of a) ladder behaves as
described in subsection 2.1, either with (top)
or without (bottom) momentum transfer along the ladder.
In this sense the
two-gluon structure function is a building block
for the four-gluon structure function presently
under study.
The contribution of the diagram in Fig.3 is 
$$\putequation{2.3.1}$$
where $F(y,r)$ is the two-gluon structure
function from subsection 2.1, and
$\tilde{F}(y,r) = {\partial \over \partial y}
{\partial \over \partial r} F(y,r)$. As under
(\puteqn{1.1}) we define $r_{l_1} = \ln(l_1^2/l_{1,0}^2)$, etc.

We will now explain the construction of the contribution in 
(\puteqn{2.3.1})
piece by piece, starting with the prefactors.
First, due to the fact that there is momentum transfer along
ladders in the top half of the diagram,
and we thus have to employ (\puteqn{2.1.6}),
we need to take into account the momentum
factor in eq. (\puteqn{2.1.6}) and carefully integrate this factor over
the angle between $l_{2t}$ and $l_{1t}$. It gives an additional
factor $1/2$ in front of the whole expression. Note that there
is no such factor for the bottom half.
Second the color factor in front
of the diagram of Fig.3 is smaller than
for the diagram of Fig.2, by a factor of $1/(N^2_c-1)$,
see Figs.4a and 4b.
Third, there is a factor of 2 from the bottom half
due to the combinatorics of connecting the rungs to 
the `vertical lines'. 
Fourth, the explicit factor of $\alb$ means that even
at lowest order, we have 1 rung for this contribution.
Not having this would mean that the lowest order
corresponds to the lowest order for the 
two-ladder exchange contribution
discussed in the previous subsection. Thus we
avoid double counting.

We remark that, because we work in DLA, besides the momenta, also
all rapidities are strongly ordered and decrease
from top to bottom along ladders.
The factors on the first line of (\puteqn{2.1.6})
correspond to the two ladders in the top half of
Fig.3. The transverse momenta here range from 
$Q^2$ down to $l_{1t}^2$, the 
momenta at the boundary of the top and bottom half.
The expression in curly brackets
corresponds to the bottom half of Fig.3, the two terms
within these brackets 
corresponding to the case $y_2 < y_1$, and the second
to $y_2 > y_1$
(we have interchanged integration variables
$y_2$ and $y_1$ for this term).
Eq. (\puteqn{2.3.1}) then is the contribution of Fig.3 summed
over all possible values of $y_1$ and $l_1^2$.

We now need to perform the integrals in (\puteqn{2.3.1}).
This is discussed in the appendix. The final answer is
$$\putequation{2.3.2}$$
where $\delta=1/(N_c^2-1)$. Next, we will add the contributions
from the previous two subsections to all orders in the 
Pomeron coupling.

\subsection{2.4. {\it The value of the anomalous dimension.}}

To get the value of the anomalous dimension we need to sum up
all diagrams with Pomeron-Pomeron interaction. For two Pomerons
is it easy to get the answer, due to the fact that we only need
to sum all diagrams of Fig.5. They form a geometric series,
the first two terms of which are given in (\puteqn{2.2.2}) and
(\puteqn{2.3.2}).
Using this, the sum can be written in the form 
$$\putequation{2.4.1}$$
from which we see that there is a pole to the right
of $\omega f = 4 \alb$, namely at
$$\putequation{2.4.2}$$
Near this singularity the Green function has the form
$$\putequation{2.4.3}$$
Directly from eq. (\puteqn{2.4.3}) we can get the value of the anomalous
dimension
$$\putequation{2.4.4}$$
The rather complicated form of the term in brackets in eq.(\puteqn{2.4.1})
originated in the sum of two terms in eq.(\puteqn{2.3.1}). However, none
of these complications are essential in the vicinity
of the rightmost pole.
Indeed we can consider both terms in eq. (\puteqn{2.3.1}) as equal and write
down the contributions of the diagram in Fig.3 in the following way
$$\putequation{2.4.5}$$
where $\lambda=4\alb/(N_c^2-1) = 4 \alb \delta$. 
This expression has a transparent meaning, namely,
it describes the interaction between
two particles with the Green function of eq.(\puteqn{2.1.5}).
This interaction is attractive, even though $\lambda > 0$.
This is explained in the next section, under eq.(\puteqn{prop2}).
The sum of the diagrams of Fig.5 gives then
$$\putequation{2.4.6}$$
It is easy to check that in the vicinity of 
$\omega f = 4 \alb(1+ \delta^2)$ this equation gives the
same pole as eq.(\puteqn{2.4.3}).

{\bfs\section{The effective two-dimensional theory.}}

To calculate the anomalous dimensions of higher twist operators
we develop here an approach via an effective two-dimensional
theory, noting that the rescattering of Pomerons does not
change the number of Pomerons (see Fig.6). It means that in 
fact we are dealing with a quantum mechanical theory. To
specify it, let us note that in $(\om,f)$ representation
the `Pomeron' propagator of eqn. (\puteqn{2.1.5}) can be
treated as the propagator of a two-dimensional particle
which is written in light-cone variables. The variables
$\om$ and $f$ play the role of momenta $k_+$ and $k_-$,
and $\sqrt{\alb}$ the role of mass. The total `momentum'
is conserved in interactions. As we showed above, the 
leading two-Pomeron pole is located near the value of 
the branchpoint $4\alb$, whereas the Pomeron-Pomeron
interaction $\lambda = 4\alb\delta$ can be considered
small compared to the Pomeron mass. This will enable us
to use a nonrelativistic approach to the n-Pomeron
interaction problem.

To specify this theory let us consider an arbitrary diagram
with $n$ Pomerons in the $t$-channel. In the $\om,f$ representation
the propagator for $n$ Pomerons can be written as
$$\putequation{3.1}$$
The sum $\sum_k^n f_k =f $ is conserved throughout all diagrams.
Eq. (\puteqn{3.1}) can be obtained by integrating eq. (\puteqn{2.1.5}) for 
each Pomeron in the $t$-channel separately over $\om_k$ 
($\om_k = \om_{k,0} = {4\alb\over f_k}$). 
Now let us introduce a new variable $\Delta_i$ such that
$$\putequation{3.2}$$
The expansion around $f/n$ looks very natural since we
believe that, due to the smallness of the interaction
$\lambda = 4\alb/(N_c^2 -1)$ at large $N_c$ the dominant
contribution to $f_k$ in the integral comes still 
from the saddle point approximation value $f_k = f/n$
(\puteqn{saddlef2}). 
Assuming that $\Delta_k << 1$ and taking into account that
$\sum_k^n \Delta_k = 0$ we can get instead of (\puteqn{3.1})
the following expression for the propagator
$$\putequation{prop2}$$
Let us interpret this propagator as $(H-{\cal E})^{-1}$
where H is some nonrelativistic Hamiltonian $p^2/2m$ 
and ${\cal E}_n$ one of its eigenvalues.
Here ${\cal E}_n = -\om +(\alb/f)n^2$ and $m= f/2\alb n$.
One should note, however the essential difference between the
usual quantum mechanical system and the Pomeron one. Due
to opposite sign of $\om$ and ${\cal E}_n$,
the pole
corresponding to the bound state of n-particles will be not
to the left, as is usual, but to the right of the multiparticle-threshold
branchcut, with branchpoint $n\alb$.
This sign is also the reason that in (\puteqn{2.4.6}) $\lambda > 0$
corresponds to an attractive interaction.
As explained in section 2, we need the rightmost singularity
in $f$, which translates into the lowest value of ${\cal E}_n$. Thus,
our task is to formulate the theory specified by $H$ and
determine its groundstate energy, the energy of its n-particle bound state.

We can now write down for the $n$-Pomeron system the following
Hamiltonian
$$\putequation{Hammom}$$
where $a^{\dagger}_k, a_k$ are bosonic creation- and annihilation
operators. 
This corresponds to the coordinate space form
$\int dx \,{1\over 2m} \partial_x \psi^{\dagger}\partial_x \psi
+ \lambda : \psi^{\dagger}\psi^{\dagger}\psi\psi:$,
or, after rescaling $x$:
$$\putequation{Hamcoo}$$

{\bfs\section{Solution with Bethe Ansatz.}}

The Hamiltonian (\puteqn{Hamcoo}) is well known, and methods
have been developed to find its spectrum (\putref{BIK}), for 
certain boundary conditions.
Our task is to find the energy of the ground state for the $n$-Pomeron
system in the $t$-channel, which corresponds to finding
the simultaneous eigenfunction and eigenvalue of the Hamiltonian
in (\puteqn{Hamcoo}), and the number operator
$$\putequation{numbop}$$
In other words, we would like to solve
$$\putequation{eigen}$$
The standard procedure of solution is given in [\putref{BIK}],
but we repeat here the outline for completeness.

The Bethe ansatz method consists of parametrizing the
eigenfunction as 
$$\putequation{eigfun}$$
where $|0\rangle$ is the Fock vacuum defined by
$$\putequation{Fockdef}$$
For the function $\chi_n$, eq. (\puteqn{eigen}) can be 
rewritten in the form
$$\putequation{eigenz}$$
where
$$\putequation{Hamz}$$
Eq. (\puteqn{eigenz}) can be solved now as follows.
From (\puteqn{Hamz}) we see that for the sector
$z_1 < z_2 < \ldots < z_n$ $\chi_n$ is an eigenfunction
of the free Hamiltonian
$$\putequation{Hamfree}$$
The interaction term in (\puteqn{Hamz}) contributes only
to the boundary conditions
$$\putequation{bound}$$
The eigenfunctions in (\puteqn{Hamfree}) are obvious:
$$\putequation{chi0}$$
with eigenvalue
$$\putequation{enfree}$$
The eigenfunction $\chi_n$ has been found in [\putref{BIK}]
and is 
$$\putequation{chiint}$$
This $\chi_n$ satisfies both (\puteqn{Hamfree}) and
(\puteqn{bound}). (see [\putref{BIK}] for details). 
The determinant can be written as a sum over all permutations
${\rm P}$ of the number $1\ldots n$:
$$\putequation{detP}$$
where $[{\rm P}]$ is the parity of the permutation. Substituting
(\puteqn{detP}) in (\puteqn{chiint}) and performing all derivative
operations one obtains
$$\putequation{chifin}$$
where $\varepsilon(\kappa)$ is the sign-function.
To determine the values of $\kappa_i$ we have to invoke
a boundary condition for our $n$-particle system. We
assume that the ground state for this system occurs
when the $n$-particle wave function falls down exponentially
when any $z_i \rightarrow \infty$ (i.e. for the n-particle
bound state), with the additional 
condition
$$\putequation{sumlam}$$
which means that we are in the center of mass of the $n$-particle
system. The solution was already found in [\putref{McG}] and is simply:
$$\putequation{lambdas}$$
Substituting (\puteqn{lambdas}) in (\puteqn{enfree})
we can calculate
$$\putequation{enint}$$
So finally the value of the anomalous dimension for the
twist $2n$ operator is 
$$\putequation{gamma2n}$$
where $\delta = (N_c^2-1)^{-1}$. This is our main result.

Let us illustrate (\puteqn{lambdas})
by giving two examples.
First, we discuss the two-Pomeron case (i.e. the anomalous
dimension of the twist four operator).
Since we have already
considered this case separately, it is instructive to see if
we obtain the same result as in section 2 from the Bethe
ansatz. The eigenfunction (\puteqn{chifin}) for the case
$n=2$ is equal to
$$\putequation{chin=2}$$
Using the constraint $\kappa_1+\kappa_2=0$, we can write
$$\putequation{chin=2p}$$
From this equation we derive that $\kappa_1-\kappa_2 =
2\kappa_1 = -i\lamb$ if we demand exponential fall-off
of the wave function as $z_2-z_1 \rightarrow \infty$.
Substituting $2\kappa_1 = -i\lamb$ in (\puteqn{chin=2})
and normalizing the wavefunction we find
$$\putequation{chin=2fin}$$
With (\puteqn{enint}) and the prescription below (\puteqn{prop2})
we recover our old result for $\gamma_4(\om)$
$$\putequation{gamma4p}$$

Next we discuss the three-Pomeron case (i.e. the anomalous
dimension of the twist six operator).
We can write the wavefunction $\chi_3(z_1,z_2,z_3)$
according to (\puteqn{chifin}), as follows
$$\putequation{chin=3}$$
where $\kappa_{ij} = \kappa_i-\kappa_j$.
Here we have factored out the center of mass coordinate 
$r=(z_1+z_2+z_3)/3$, ordered the imaginary parts of the $\kappa$'s 
($\Im\kappa_3>\Im\kappa_2>\Im\kappa_1$) and used 
the relative coordinates
$x_i = z_i-r,\, i=1,2,3$. We also employed the 
relation $\kappa_1+\kappa_2+\kappa_3=0$.

From this example and the previous one we can easily understand why 
(\puteqn{lambdas}) is the solution. Indeed, let us
assume that $x_3 >> x_2 $. 
The terms with  a plus sign for the $x_3$ give a rising exponent.
Thus the coefficient in front of such a term must be zero.
The solution (\puteqn{lambdas}) for the case $n=3$ 
accomplishes just that. The wavefunction
for this case is then
$$\putequation{chi3bound}$$
The above generalizes  to arbitrary $n$ [\putref{McG}].
We conclude this section
with the observation of a remarkable feature of the solution
of the $n$-Pomeron exchange amplitude: 
although we started out with a completely bosonic system,
we find a non-degenerate momentum spectrum, implying
that in this sense the Pomerons behave as fermions.
Fig.7 shows the one particle levels in our
system of interacting bosons (Pomerons) and it
is easy to see that the direction of motion plays
a role in the spin of the fermion since each level
has two Pomerons moving in different directions.
We believe that the understanding of this property
is another important result of this paper. We will discuss it further
in the conclusions.

{\bfs\section{Conclusions.}}

Let us repeat our main assumptions 
which led us to the value of the anomalous dimension
of eq. (\puteqn{gamma2n}). 

1) We assumed that only Pomeron-Pomeron interactions
contribute to the value of $\gamma_{2n}$. The experience
from the exact solution of the next-to-leading twist
anomalous dimension taught us that other color states
in the t-channel for four gluons, due to diagrams such as
depicted in Fig.8, lead simply to 
a renormalization of $\lambda$.
Such a renormalization can
easily be accounted for by replacing $\lambda$
with $\tilde{\lambda}$, taking the latter from [\putref{B,LRS}].

2) The problem of the contribution of color states in the 
system of 6 or more gluons, which would induce a direct
interaction between 3 and more Pomerons (see Fig.9) is still
an open one.

3)We would like to stress again that we can trust the
answer of (\puteqn{gamma2n}) only if $\delta n^2/3 << 1$, 
which reflects our assumption that $\Delta_k << 1$.

Our calculated value of $\gamma_{2n}$ leads to the following
new insights.

Firstly, we cannot trust the GLAP evolution equation in the 
region of small $x_B$ (or $\om\rightarrow 0$), since the
high-twist contributions rapidly become more important in the Wilson OPE
than the leading twist one. Recall
that the GLAP evolution equation can be used only to calculate
the anomalous dimension of the leading twist contribution.

Secondly, to obtain the correct evolution equation in the region
of small $x_B$ we need to sum all high-twist contributions in the 
Wilson OPE. The nonlinear GLR equation is an example of an
evolution equation that does take into account all high-twist
contributions, but it should be noted that this equation
assumed  for the anomalous dimensions the
simple value $\gamma_{2n} = \alb n^2/\om$.
Thus the result of the present paper shows that the GLR
equation cannot be valid and could only have some numerical
accuracy related to small values of $\delta^2$.
In other words, only in the limit
$N_c \rightarrow \infty$ does the GLR equation correctly
take into account the high-twist contributions 
which induce shadowing corrections.

Thirdly, the calculated value of $\gamma_{2n}$ shows us 
that the theory with a large number of colors ($N_c \rightarrow \infty$)
cannot be a good approximation to reality, since the high-twist
contributions with $n>N_c$ are larger than the ones with
$n<N_c$.

Finally, we found a remarkable regularity in
the bosonic system of $2n$ gluons inside of the parton
cascade: they behave in a certain sense as fermions.
The direction of motion plays the role of spin here,
and obeys Fermi statistics. The above property is
the direct consequence of the two-dimensional character
of our DLA theory. The nonrelativistic, fermionic nature
of our system allows us to understand why the bosonic
system does not collapse, in spite of the attractive
forces acting betweens the bosons (Pomerons). This
result, perhaps, indicates a way to a more systematic statistical
description of the parton cascade in a situation
with a large parton density. 

Some open problems spawned by these considerations are the following.
First, it should be understood what the restricted kinematical
region is where the nonlinear GLR equation is sufficiently accurate
in estimating the value of shadowing corrections. Second,
we need a generalization of the GLR equation which includes the
correct behavior of $\gamma_{2n}$ when $\om \rightarrow 0$.
And finally, the calculated value
of $\gamma_{2n}$ shows that deviations from the GLAP
equation in deep-inelastic scattering and other hard hadronic
processes should enter earlier than was estimated
in the framework of the GLR equation. This is perhaps encouraging
in the search for experimental signals of shadowing corrections.

\sectionnumstyle{blank}

{\bfs\section{Appendix A}}

\vskip0pt plus.5\baselineskip

\sectionnumstyle{Alphabetic}
\sectionnum=1
\equationnum=0

In this appendix we derive (\puteqn{2.3.2}), starting
from (\puteqn{2.3.1}). We start by transforming this
expression to $\om,f$ space by applying
$\int_{-\infty}^{\infty} dY \exp(-\om Y)
\int_{-\infty}^{\infty} dr_Q \exp(-f r_Q)$. 
Since the complete integral
vanishes for negative $Y$ and $r_Q$,
we have extended the integrals down to $-\infty$ 
to facilitate our calculations.
We now focus 
on the overall factor $F(Y-y_1,r_Q-r_{l_1})^2$ first,
and use the representation
$$\putequation{App1}$$
with $G_2$ given in (\puteqn{2.1.5}). Performing
the $Y$ and $r_Q$ integrals (after a change of variables)
we obtain, similarly to (\puteqn{2.2.1}) ,(\puteqn{2.2.2}), the
factor
$$\putequation{App2}$$
The remainder then takes the form
$$\putequation{App3}$$
Performing the integrals over $y_1$ and $r_{l_1}$ yields
a product of $\delta$-functions: $\delta(\om - \om_1 - \om_2)
\delta(f - f_1 - f_2)$. Next one performs the $\om_2$ integral
by closing around the pole $\alb/f_2$. Then using the $\delta$-functions
for the $\om_1$ and $f_1$ integrations one is left with
$$\putequation{App4}$$
This integrand has poles in $f_2$ at 
$f_2^0 = \alb/\om$ and $f_2^{\pm} = f(1\pm \sqrt{1-{4\alb\over \om f}})/2$.
The contour of $f_2$ runs to the right of $f_2^0$, but we can close
it either to the left or to the right, since there is no longer
an exponent to guide us. It is easy to see that in the limit
of large $\om$ the pole $f_2^-$ migrates to $f_2^0$, so 
that we best close the contour on to the right, picking 
up only the $f_2^+$ pole. After some straightforward algebra
we find that the result for the integral in 
(\puteqn{App4}) is 
$$\putequation{App5}$$
Combining this with the result for the overall factor in 
(\puteqn{App2}), and multiplying by 2 to take into
account the case $l_2^2 >> l_1^2$, we arrive at
(\puteqn{2.3.2}).

\sectionnumstyle{blank}

{\bfs\section{References}}

\footnotesize
\raggedbottom

\begin{putreferences}

\reference{Coll}{J.C. Collins, {\it Renormalization}, Cambridge
	University Press, Cambridge, 1984}

\reference{GL}{V.N. Gribov and L.N. Lipatov, Sov. J. of Nucl. Phys. {\bfs 15}
	(1972) 438; \\
	L.N. Lipatov, Yad. Fiz. {\bfs 20} (1974) 181; \\
	 G. Altarelli, G. Parisi, Nucl. Phys. {\bfs B126} (1977) 298.}

\reference{GLR}{L.V. Gribov, E.M. Levin and M.G. Ryskin, Phys. Rep. {\bf 100}
	(1983) 1.}

\reference{B}{J. Bartels, Phys. Lett. {\bfs B298} (1993) 204.}

\reference{LRS}{L.V. Gribov, E.M. Levin and A.G. Shuvaev, 
 Nucl. Phys. {\bfs B387} (1992) 589}

\reference{BIK}{N.M. Bogoliubov, A.G. Izergin, V.E. Korepin,
{\it Lecture Notes in Physics}, Vol. 242, Springer Verlag, (1986)
220.}

\reference{MW}{B.M. McCoy and T.T. Wu}

\reference{MQ}{A.H. Mueller and J. Qiu}

\reference{MN}{A.H. Mueller and H. Navelet}

\reference{McG}{J.B. McGuire, J.Math.Phys. {\bfs 5} (1964) 622.}

\reference{EFP}{R.K. Ellis, W. Furmanski and R. Petronzio,
	Nucl. Phys. {\bfs B207} (1982) 1; {\bfs B212} (1983) 29.}

\reference{BFL}{A.P. Bukhvostov, G.V. Frolov and L.N. Lipatov,
 Nucl. Phys. {\bfs B258} (1985) 601.}

\end{putreferences}

\newpage

	\centertext{\bf FIGURE CAPTIONS}

\typesize=12pt
\parindent=0pt

{\bf Figure 1.}

The two-gluon structure function $F(x,q_1^2, q_2^2)$. $q_{1t}$
and $q_{2t}$ are the transverse components of the gluon momenta
$q_1$ and $q_2$, whereas $q_{0,1t}$ and $q_{0,2t}$ are cut-off
momenta. 

\smallskip

{\bf Figure 2.}

The two-ladder contribution. $Q^2$ is the photon mass, $q_0^2$ is
an IR cut-off. $q_t$ is the transverse momentum along the ladder. 

\smallskip

{\bf Figure 3.}

A twist-four contribution to the gluon structure function.
Here $k_i$, $m_i$, $q_t$, $l_1$ and $l_2$ are all transverse momenta,
all `y' 's are rapidities, and the IR cut-off momenta $l_{0,1}$,
$l_{0,2}$ are both of order $1/R_h$.

\smallskip

{\bf Figure 4.}

Fig. 4b shows that at the same order in $\alpha_S$ the two-ladder
contribution has one power of $N_c^2-1$ more than the Pomeron
interaction contribution in Fig. 4a. Again, the Pomerons are
represented by gluon ladders. The wiggly lines also represent 
gluons.

\smallskip

{\bf Figure 5.}

A ladder of Pomeron-interactions, whose contribution is
given in (\puteqn{2.4.1}).

\smallskip

{\bf Figure 6.}

Pomeron interactions. Note the total number of Pomerons is
always constant. 

\smallskip

{\bf Figure 7.}

One particle levels for Pomerons in the t-channel for 
fixed n 
(here n=8). Note that there are two states per energy level,
and that thus the direction of motion behaves as a spin
quantum number.

\smallskip

{\bf Figure 8.}

Interaction of non-singlet ladders. This leads to a renormalization
of the coupling $\lambda$.

\smallskip

{\bf Figure 9.}

Three Pomeron interaction. This has not been taking into account
in this paper.

\bye